\documentclass[12pt]{iopart}

\usepackage{graphicx}
\usepackage{bm}
\usepackage{amsbsy}
\usepackage{amssymb}
\usepackage{mathrsfs}

\newcommand{\ave}[1]{\langle #1 \rangle}
\newcommand{\bra}[1]{\langle #1|}
\newcommand{\ket}[1]{| #1 \rangle }

\newcommand{\cE}{{\cal E}}

\begin{document}

\title[Single photon-added coherent states: estimation of parameters]{Single photon-added coherent states: estimation of parameters and fidelity of the optical homodyne detection}

\author{S~N~Filippov$^{1,2}$, V~I~Man'ko$^{1,3}$, A~S~Coelho$^4$, A~Zavatta$^{5,6}$, M~Bellini$^{5,6}$}

\address{$^1$ Moscow Institute of Physics and Technology, 141700 Moscow Region, Russia}

\address{$^2$ Institute of Physics and Technology, Russian Academy of Sciences, 117218 Moscow, Russia}

\address{$^3$ P N Lebedev Physical Institute, Russian Academy of Sciences, 119991 Moscow, Russia}

\address{$^4$ Instituto de F\'{i}sica, Universidade de S\~{a}o Paulo, 05315-970 S\~{a}o Paulo, Brazil}

\address{$^5$ Istituto Nazionale di Ottica, INO-CNR, L.go E. Fermi, 6, I-50125 Florence, Italy}

\address{$^6$ LENS, Via Nello Carrara 1, I-50019 Sesto Fiorentino, Florence, Italy}

\eads{\mailto{sergey.filippov@phystech.edu}}

\begin{abstract}
Travelling modes of single-photon-added coherent states (SPACS)
are characterized via optical homodyne tomography. Given a set of
experimentally measured quadrature distributions, we estimate
parameters of the state and also extract information about the
detector efficiency. The method used is a minimal distance
estimation between theoretical and experimental quantities, which
additionally allows to evaluate the precision of estimated
parameters. Given experimental data, we also estimate the lower
and upper bounds on fidelity. The results are believed to
encourage preciser engineering and detection of SPACS.\\[-1cm]
\end{abstract}



\section{\label{introduction}Introduction}
Optical homodyne tomography is a powerful technique to infer
continuous-variable quantum states of a specific mode of
electromagnetic radiation. Its history saw a dramatic boom in the
last two decades, when both theoretical and experimental methods
evolved significantly from the first proof-of-principle
studies~\cite{vogel-risken-1989,raymer,mlynek} to the
state-of-the-art detection of arbitrarily shaped ultrashort
quantum light states~\cite{polycarpou} and the experimental
analysis of decoherence in continuous-variable bipartite
systems~\cite{porzio12}. Different stages of the research in this
area can be seen in the monographs and
reviews~\cite{leonhardt-book,bachor-book,zavatta-LPL-2006,vogel-2006,walls,lvovsky},
where different approaches to the state reconstruction are
outlined and corresponding experimental realizations are
discussed.

The goal of this paper is to consider both theoretically and
experimentally the detection of single-photon-added coherent
states (SPACS). These states are defined by formula
$a^{\dag}\ket{\alpha} / \sqrt{1+|\alpha|^2}$, where $\ket{\alpha}$
is a conventional coherent state ($\alpha\in\mathbb{C}$) and
$a^{\dag}$ is a photon creation operator. Photon-added states of
light and their non-classical properties were considered
originally in the papers~\cite{agarwal,dodonov-manko-kor-mukh} and
then realized in
practice~\cite{zavatta-science-2004,zavatta-pra-2005}. The
techniques of photon addition and subtraction allowed to check
experimentally the commutation relation between the corresponding
operators~\cite{parigi-science-2007,kim-jeong-zavatta-prl-2008,zavatta-prl-2009}.
The process tomography of photon creation and annihilation
operators was recently reported~\cite{kumar}. Nonclassical
behaviour of photon-added states was demonstrated in the
papers~\cite{zavatta-pra-2007,parigi-jpa-2009,kiesel-2011} and
noiseless amplification was discussed in
Ref.~\cite{zavatta-natphot-2011}.

The practical homodyne detection of some signal results in
experimental quadrature distributions $w_{\rm ex}(X,\theta)$ to be
compared with theoretically predicted ones $w_{\rm th}(X,\theta)$.
The adequate theoretical model should take losses into account,
which are usually modelled by fictitious beamsplitters with
transmittivity $\eta$ placed in front of ideal detectors. In the
paper~\cite{zavatta-pra-2005}, the explicit form of such
theoretical quadrature distributions $w_{\rm th}(X,\theta)$ for
SPACS is found for any $\eta$. We can associate density operators
$\rho_{\rm ex}$ and $\rho_{\rm th}$ with distributions $w_{\rm
ex}(X,\theta)$ and $w_{\rm th}(X,\theta)$, respectively. Note that
these states are mixed in general and depend on parameters
$\alpha$ and $\eta$.

We can naturally define the fidelity of detection as fidelity
between $\rho_{\rm ex}$ and $\rho_{\rm th}$, i.e. $F = {\rm Tr}
|\sqrt{\rho_{\rm ex}} \sqrt{\rho_{\rm th}}| \equiv {\rm Tr} \sqrt{
\sqrt{\rho_{\rm th}} \rho_{\rm ex} \sqrt{\rho_{\rm th}} }$. It is
tempting to express $F$ directly through the measured
distributions $w_{\rm ex}(X,\theta)$ avoiding reconstruction of
the state $\rho_{\rm ex}$ and dealing with complicated formulas.
The easiest way is to find the Bhattacharyya
coefficient~\cite{bhattacharyya} for distributions $w_{\rm
ex}(X,\theta)$ and $w_{\rm th}(X,\theta)$, which turns out to be
the upper bound for $F$~\cite{filippov-manko-distances}.
Alternatively, one can use upper and lower bounds for $F^2$
developed in the paper~\cite{uhlmann-2008} and also known as
super- and sub-fidelity, respectively. In this paper, we present
operational ways to calculate these quantities.

In principle, maximizing the sub-fidelity with respect to $\alpha$
and $\eta$ would enable us to estimate both these parameters. As
it will be shown by an example in \Sref{section-fidelity}, such a
method can be applied to extremely precise data only. If this is
not the case, parameters $\alpha$ and $\eta$ can be estimated by
minimizing another distance between the states $\rho_{\rm ex}$ and
$\rho_{\rm th}$ (not the Bures distance related to the fidelity).
Fortunately, the Hilbert--Schmidt distance is easy to compute via
tomograms and its minimization is performed in
\Sref{section-estimation}. As a result, an operational estimation
of state and measurement parameters is achieved. Finally, errors
of the estimated parameters are evaluated by using symmetry
condition $w(X,\theta+\pi) = w(-X,\theta)$ met by fair optical
tomograms. This approach was suggested and demonstrated in the
papers~\cite{filippov-fss-11} and~\cite{bellini-2012},
respectively. The improved precision of homodyne detection is of
vital importance to check different uncertainty relations
(see~\cite{bellini-2012} and references therein) as well as to
probe commutation relations between position and momentum of
massive particles, which may be modified by gravity and feasibly
detected with the help of quantum optics~\cite{pikovski}.

The paper is organized as follows.

In \Sref{section-quadrature-distributions}, we remind the explicit
formula of homodyne quadrature distributions of SPACS modified by
the losses. In \Sref{section-estimation}, we present theoretical
basics and demonstrate particular results of the minimal distance
estimation of the state and apparatus parameters. In
\Sref{section-fidelity}, the fidelity of detection is discussed.
In \Sref{section-conclusions}, we briefly resume the results
obtained and outline the prospects.

\section{\label{section-quadrature-distributions} Quadrature distributions of SPACS}

Generation of SPACS is due to injection of a coherent state
$\ket{\alpha}$ into the signal mode of an optical parametric
amplifier. The stimulated emission of a single down-converted
photon into the signal mode results in SPACS generation, which is
trigged by the detection of a single photon in the idler mode of
the amplifier. A time-domain balanced homodyne detector is then
used to acquire quadrature data (see, e.g., the
review~\cite{zavatta-LPL-2006}).

The balanced homodyne detection is known to give access to
quadratures $\hat{X}_{\theta} = \hat{Q}\cos\theta + \hat{P}
\sin\theta$, where $[\hat{Q},\hat{P}] = \rmi$ and $\theta \in
[0,2\pi]$ is a phase of an intense coherent light (the local
oscillator). Once $\theta$ is fixed, the distribution of
quadratures is given by the optical tomogram $w_{\rm th}(X,\theta)
= \langle X_{\theta} | \rho | X_{\theta} \rangle$, where $\rho$ is
the density operator of quantum state and $\hat{X}_{\theta} |
X_{\theta} \rangle = X | X_{\theta} \rangle$.

Let $\rho$ be a density operator of SPACS, then the tomogram
$\widetilde{w}_{\rm th}(X,\theta) = \langle X_{\theta} | \rho |
X_{\theta} \rangle$ is easy to compute. However, it turns out that
the experimentally measured quadrature distributions are smoother
than the predicted ones and can differ significantly from them.
This takes place due to losses and overall efficiency of detection
$\eta < 1$. One can make allowance for losses by introducing a
fictitious beamsplitter with transmittivity $\eta$ in front of the
ideal photodetectors (with sensitivity of 100\%). Such an
attenuation of the signal results in the following convolution
relation between the quadrature
distributions~\cite{leonhardt-paul-1993}:
\begin{equation}
\label{convolution} \!\!\!\!\!\!\! w_{\rm th}(X,\theta;\eta) =
\frac{1}{\sqrt{\pi(1-\eta)}} \int \widetilde{w}_{\rm th}(Y,\theta)
\exp\left[ -\frac{\eta}{1-\eta} \left( Y - \sqrt{\eta}X \right)^2
\right] dY.
\end{equation}

In the Schr\"{o}dinger picture, the distribution $w_{\rm
th}(X,\theta;\eta)$ is nothing else but the optical tomogram
$\langle X_{\theta} | \cE_{\eta}[\rho] | X_{\theta} \rangle$ of
the transformed state $\cE_{\eta}[\rho]$, where
$\cE_{\eta}[\bullet] = \sum_{k=0}^{\infty} A_k(\eta) \bullet
A_k^{\dag}(\eta)$ is a completely positive trace preserving map
with the following operator-sum representation: $A_k(\eta) =
\sum_{m=0}^{\infty} \sqrt{\frac{(m+k)!}{m!k!} \, \eta^m
(1-\eta)^k} \, \ket{m}\bra{m+k}$ (see,
e.g.,~\cite{leonhardt-book,ivan}).

Using formula (\ref{convolution}), one can calculate in explicit
form the optical homodyne tomogram of a SPACS. Some algebra yields
\begin{eqnarray}
\label{w-alpha-eta} \!\!\!\!\!\!\!\!\!\! w_{\rm
th}(X,\theta;\alpha,\eta) &=& \frac{1}{\sqrt{\pi}(1+|\alpha|^2)}
\Bigg\{ (1-\eta)
\Big( 1+4\eta |\alpha|^2 \sin^2(\theta-\varphi) \Big) \nonumber\\
&& +2\eta \Big[ \Big( X\cos(\theta-\varphi) - \frac{2\eta -
1}{\sqrt{2\eta}} |\alpha| \Big)^2 + X^2 \sin^2(\theta-\varphi)
\Big] \Bigg\} \nonumber\\
&& \times \exp \Big[ - \left( X-\sqrt{2\eta}
\,|\alpha|\cos(\theta-\varphi) \right)^2 \Big].
\end{eqnarray}

\noindent An analogue of tomogram (\ref{w-alpha-eta}) was first
derived in the paper~\cite{zavatta-pra-2005}, where the authors
used a slightly different commutation relation $[\hat{Q},\hat{P}]
= \frac{\rmi}{2}$. The deduced tomogram (\ref{w-alpha-eta})
comprises two parameters: $\alpha=|\alpha|\rme^{\rmi\varphi}$
determines the coherent state $\ket{\alpha}$ to which a single
photon is added, $\eta$ is the overall efficiency of homodyne
detection and characterizes the imperfection of measurement
device. The overall efficiency includes transmission losses, mode
matching and the intrinsic quantum efficiency of detectors.

\section{\label{section-estimation} Estimation of parameters}

Our goal is to compare $w_{\rm th}(X,\theta;\alpha,\eta)$ with the
experimentally measured distributions $w_{\rm ex}(X,\theta)$ and
find parameters $\alpha=|\alpha|\rme^{\rmi\phi}$ and $\eta$
resulting in the best fitting. In this sense, we perform a minimal
distance estimation of the state parameter $\alpha$ and the
detector parameter $\eta$. In order to give this procedure more
rigorous formulation with clearer physical meaning, we need to
choose such a distance between distributions $w_{\rm
th}(X,\theta;\alpha,\eta)$ and $w_{\rm ex}(X,\theta)$ that is
related with some fair distance between states $\rho_{\rm th}
\equiv \cE_{\eta}[\rho]$ and $\rho_{\rm ex}$ (satisfying metric
requirements). Moreover, we are interested in such a distance
between the states that could be operationally calculated via
optical tomograms. Some aspects of appropriate distances were
discussed in the paper~\cite{dodonov-manko-wunsche-1999}. The
Hilbert--Schmidt distance $D = \sqrt{\Tr(\rho_{\rm th} - \rho_{\rm
ex})^2}$ turns out to be suitable because it can be given by the
following expression in terms of tomograms:
\begin{eqnarray}
\label{D} &&
\!\!\!\!\!\!\!\!\!\!\!\!\!\!\!\!\!\!\!\!\!\!\!\!\!\!\!\!\!\!\!\!\!\!
D^2(\alpha,\eta) = \frac{1}{\pi} \int_{0}^{+\infty} d r ~ r
\int\!\!\!\int_{-\infty}^{+\infty} d X d Y \, \cos[(X+Y)r]
\nonumber\\
&& \!\!\!\!\!\! \times \int_{0}^{\pi} d \theta \, \Big[ w_{\rm
ex}(X,\theta) - w_{\rm th}(X,\theta;\alpha,\eta) \Big] \Big[
w_{\rm ex}(-Y,\theta) - w_{\rm th}(-Y,\theta;\alpha,\eta) \Big],
\end{eqnarray}

\noindent which can be readily deduced with the help of a formula
for $\Tr{\rho_1\rho_2}$ obtained in Ref.~\cite{manko-elaf-11}.
Similarly, the experimental error is evaluated by a slight
modification of formulas in the paper~\cite{bellini-2012}, namely,
\begin{eqnarray}
\label{delta-D} && \Delta(D^2) = \frac{1}{2\pi} \int_{0}^{+\infty}
dr ~ r \int\!\!\!\int_{-\infty}^{+\infty} d X d Y
\cos[(X+Y)r] \nonumber\\
&& \times \int_{0}^{\pi} d \theta \Big[ w_{\rm ex}(X,\theta)
w_{\rm ex}(-Y,\theta) - w_{\rm ex}(X,\theta+\pi) w_{\rm
ex}(-Y, \theta+\pi) \nonumber\\
&& \qquad\qquad + 2 w_{\rm th}(X,\theta) \Big( w_{\rm
ex}(Y,\theta+\pi) - w_{\rm ex}(-Y,\theta) \Big) \Big].
\end{eqnarray}

Formula (\ref{delta-D}) is based on the fact that the fair
quadrature distributions satisfy the symmetry relation
$w(X,\theta+\pi) = w(-X,\theta)$. Experimentally measured
distributions do not satisfy precisely this relation, and this
gives rise to the error (\ref{delta-D}) which includes both
systematic and statistical components (see details in the
paper~\cite{bellini-2012}).

\subsection{Results}

\begin{figure*}
\includegraphics{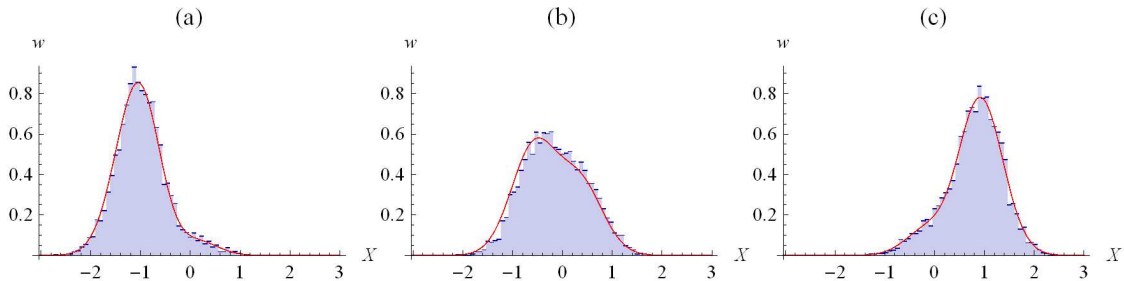}
\caption{\label{figure:histograms} Experimental histograms $w_{\rm
ex}(X,\theta)$ (blue discontinuous lines) and the closest
theoretical quadrature distributions $w_{\rm th}(X,\theta)$ (red
solid lines) of a SPACS for different phases: $\theta=0$ (a),
$\theta=1.36$ (b), $\theta=2.49$ (c).}
\end{figure*}

In this subsection, we estimate parameters $|\alpha|$, $\varphi$,
and $\eta$ for a particular set of experimental quadrature
distributions. Phases of the local oscillator take discrete values
$\{\theta_j\}_{j=1}^{21}$. For each fixed phase, the quadrature
distribution is a histogram of 5321 values, with the bin width
being chosen to guarantee the statistical confidence and prevent
the data from undersampling~\cite{bellini-2012}. Examples of
experimental histograms are depicted in
Fig.~\ref{figure:histograms}. Thus, the data are presented in
discrete form so the integrals in formulas (\ref{D}) and
(\ref{delta-D}) are calculated approximately by the trapezoid
method~\cite{korn}. The error of calculation is estimated in the
paper~\cite{bellini-2012} and is usually less than the
experimental quantity (\ref{delta-D}).

In our particular case, the minimization of the square of distance
$D^2(|\alpha|,\varphi,\eta)$ results in $D^2=0.0436$ which is
achieved at $|\alpha|_{\rm opt}=0.81$, $\varphi_{\rm opt}=3.14$,
$\eta_{\rm opt}=0.58$. On substituting these parameters in formula
(\ref{w-alpha-eta}), we can depict the closest theoretical
quadrature distributions (see Fig.~\ref{figure:histograms}).

\begin{figure*}
\includegraphics[width=16cm]{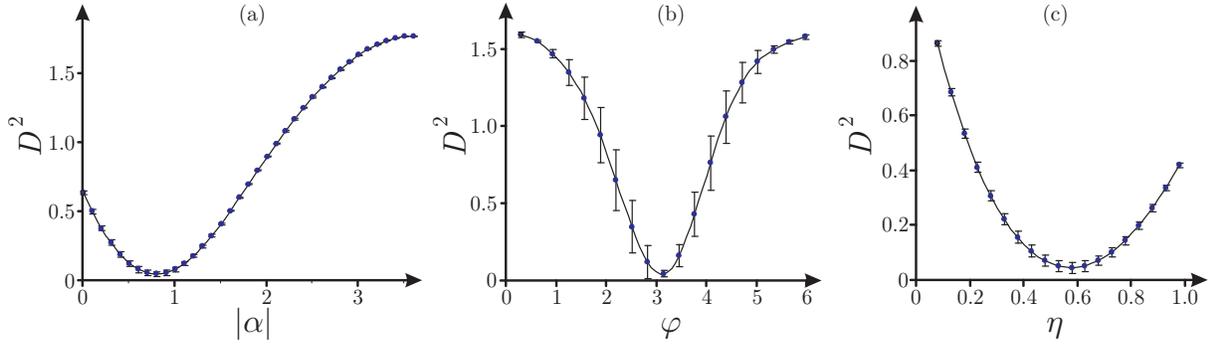}
\caption{\label{figure:alpha-eta} Square of the Hilbert--Schmidt
distance vs. state and detector parameters in the vicinity of the
global minimum: $D^2(|\alpha|,\varphi_{\rm opt},\eta_{\rm opt})$
(a), $D^2(|\alpha|_{\rm opt},\varphi,\eta_{\rm opt})$ (b),
$D^2(|\alpha|_{\rm opt},\varphi_{\rm opt},\eta)$ (c).}
\end{figure*}

In order to evaluate the errors of estimated parameters
$|\alpha|_{\rm opt}$, $\varphi_{\rm opt}$, and $\eta_{\rm opt}$ we
consider three cuts of the function $D^2(|\alpha|,\varphi,\eta)$
that cross at the point $(|\alpha|_{\rm opt},\varphi_{\rm
opt},\eta_{\rm opt})$. The values of function and their errors are
shown in Fig.~\ref{figure:alpha-eta}. Further, the error of an
optimal parameter $q_{\rm opt}$ can be evaluated as $\Delta q/{\rm
SNR}$, where $\Delta q$ is the width of the corresponding function
cut and ${\rm SNR}=(\max D^2 - \min D^2)/ \max \Delta (D^2)$ plays
the role of signal to noise ratio. The errors evaluated in such a
way give rise to the following results: $|\alpha|_{\rm opt}=0.81
\pm 0.03$, $\varphi_{\rm opt}=3.14 \pm 0.25$, $\eta_{\rm opt}=0.58
\pm 0.02$. The least precise parameter is the phase $\varphi$ and
this can be attributed to the relatively small mean number of
photons $\ave{n} \lesssim 1$ and imprecise fixing of the local
oscillator phase $\theta$. Improving control of this parameter
would result in higher precision of parameters under estimation.

\section{\label{section-fidelity} Fidelity of detection}

Sometimes, the Hilbert--Schmidt distance between the states is not
very representative because it can grow under the action of
quantum operations (not monotone metric). In this case one
exploits some other quantities, e.g., the Bures distance $D_{\rm
B}=\sqrt{2(1-F)}$, where $F = {\rm Tr} \sqrt{ \sqrt{\rho_{\rm th}}
\rho_{\rm ex} \sqrt{\rho_{\rm th}} }$ is Uhlmann's fidelity (see,
e.g., the book~\cite{Bengtsson}). The fidelity is difficult to
express in operational way through quadrature distributions.
Nevertheless, we can use recently found bounds for fidelity: the
sub-fidelity $E$ and the super-fidelity $G$ satisfying $E \le F^2
\le G$ and given by formulas~\cite{uhlmann-2008,chen,mendonca}
\begin{eqnarray}
&& E(\rho_{\rm th},\rho_{\rm ex})=\Tr\rho_{\rm th}\rho_{\rm ex} +
\sqrt{2[(\Tr\rho_{\rm th}\rho_{\rm ex})^2 - \Tr\rho_{\rm
th}\rho_{\rm ex}\rho_{\rm th}\rho_{\rm ex}]}, \\
&& \label{superfidelity} G(\rho_{\rm th},\rho_{\rm ex}) =
\Tr\rho_{\rm th}\rho_{\rm ex} + \sqrt{\left(1-\Tr\rho_{\rm
th}^2\right)\left(1-\Tr\rho_{\rm ex}^2\right)}.
\end{eqnarray}

The overlap $\Tr\rho_{\rm th}\rho_{\rm ex}$ and purity
$\Tr\rho_{\rm ex}^2$ are readily expressed through tomograms (see,
e.g.,~\cite{manko-elaf-11}). It is worth noting that the purity
can also be estimated by exploiting the covariant uncertainty
relation~\cite{porzio11}. As far as 4-product $\Tr\rho_{\rm
th}\rho_{\rm ex}\rho_{\rm th}\rho_{\rm ex}$ is concerned, we can
approximate it by $\Tr\rho_{\rm th}^4$. In fact, we have $ \vert
\Tr\rho_{\rm th}\rho_{\rm ex}\rho_{\rm th}\rho_{\rm ex} -
\Tr\rho_{\rm th}^4 \vert \le \vert \Tr\rho_{\rm ex}\rho_{\rm
th}\rho_{\rm ex} - \Tr\rho_{\rm th}^3 \vert \le \vert \Tr\rho_{\rm
ex}^2 - \Tr\rho_{\rm th}^2 \vert $ and can modify the sub-fidelity
as follows:
\begin{eqnarray}
\label{sub-fidelity-modified} \!\!\!\!\!\!\!\!\!\!\!\!
E'(\rho_{\rm th},\rho_{\rm ex}) = \Tr\rho_{\rm th}\rho_{\rm ex} +
\sqrt{2[(\Tr\rho_{\rm th}\rho_{\rm ex})^2 - \Tr\rho_{\rm th}^4 -
\vert \Tr\rho_{\rm ex}^2 - \Tr\rho_{\rm th}^2 \vert ] }.
\end{eqnarray}

In order to be able to calculate the modified sub-fidelity
(\ref{sub-fidelity-modified}) for SPACS, we find the following
theoretical values:
\begin{equation}
\!\!\!\!\!\!\!\!\!\! \Tr \rho_{\rm th}^2 =
1-\frac{2\eta(1-\eta)}{(1+|\alpha|^2)^2}, \qquad\qquad \Tr
\rho_{\rm th}^4 = 1-\frac{4\eta(1-\eta)}{(1+|\alpha|^2)^2} +
\frac{2\eta^2(1-\eta)^2}{(1+|\alpha|^2)^4}. \qquad
\end{equation}

Returning to the example considered earlier, we substitute the
experimental data and the optimal theoretical values
$|\alpha|_{\rm opt}=0.81$, $\varphi_{\rm opt}=3.14$, $\eta_{\rm
opt}=0.58$ in formula (\ref{superfidelity}) and obtain the upper
bound $G(\rho_{\rm th},\rho_{\rm ex}) = 0.98 \pm 0.02$. In our
case, the direct calculation of sub-fidelity
(\ref{sub-fidelity-modified}) turns out to be problematic because
the confidence interval of the radicand is $[-0.07; 0.05]$ (cf.
$2[(\Tr \rho_{\rm th}^2)^2-\Tr \rho_{\rm th}^4] = 0.032$). Thus,
the calculation of square root is worthless. Therefore, the use of
formula (\ref{sub-fidelity-modified}) is possible only with the
data of very high precision (errors should be substantially less
than 1\%). Whenever this does not happen, one can use another
lower bound $E''=\Tr\rho_{\rm th}\rho_{\rm ex} \le F^2$ (see,
e.g.,~\cite{uhlmann-2008}). This lower bound is easy to calculate
and in our case it equals $E''=0.81 \pm 0.02$. Consequently, the
fidelity of our interest is bounded by the two-sided inequality
$0.81 \pm 0.02 \le F^2 \le 0.98 \pm 0.02$.

\section{\label{section-conclusions}Conclusions}

In order to estimate parameters of some prepared SPACS we
developed the operational method whose essence was the comparison
of experimental histograms with theoretically predicted quadrature
distributions. The explicit form of theoretical distributions took
into account the losses presented, which allowed us to infer not
only the state parameter $\alpha$ but also the parameter $\eta$
describing the overall efficiency of homodyne detection. We
discussed some practical issues concerning the easiest way to
calculate the Hilbert--Schmidt distance and evaluate the errors of
estimated parameters. The phase of the state turned out to be the
least precise parameter, which could be ascribed to the small
intensity of the signal mode and the errors in control of the
local oscillator phase. Then we considered some operational
techniques to determine the lower and upper bounds for fidelity of
detection. We showed that, in practice, some of these bounds can
be calculated only with highly precise data.

The outlook for further research is to use high sensitivity of
homodyne detection to trace all the stages of quantum state's
life: its preparation, transformation via a quantum channel, and
detection. Using appropriate theoretical models of these
processes, one can determine the corresponding parameters. For
instance, dark counts in the trigger detector result in mixing of
the SPACS with a residual coherent state. In this case, the
measured tomogram reads $(1-p)w_{\rm SPACS} + p \, w_{\rm
coherent}$, where $p$ is a fraction of dark counts. The parameter
$p$ can be estimated by the same algorithm of comparing $w_{\rm
ex}$ and $w_{\rm th}$.

In general, optical tomograms can be valuable information sources
on equal footing with other state descriptions~\cite{ibort}.
Improving the accuracy of homodyne detection, one can check the
validity of more complicated quantum theories and observe new
phenomena (see, e.g.,~\cite{pikovski}). The role of SPACS states
for new experiments can be also dramatic because of their ability
to exhibit properties ranging from classical to quantum ones for
different intensities~\cite{zavatta-science-2004}.

\ack S.N.F. and V.I.M. are grateful to the Organizers of the 19th
Central European Workshop on Quantum Optics (Sinaia, Romania, July
2-6, 2012) for invitation and kind hospitality. S.N.F. would like
to express his gratitude to the Organizing Committee of the
Conference and especially to Dr. Aurelian Isar for financial
support. S.N.F. and V.I.M. thank the Russian Foundation for Basic
Research for partial support under projects 10-02-00312-a and
11-02-00456-a and the Ministry of Education and Science of the
Russian Federation for partial support under project 2.1759.2011.
S.N.F. also appreciates supports from the Russian Foundation for
Basic Research under project 12-02-31524-mol-a and the Dynasty
Foundation (www.dynastyfdn.com). A.S.C. acknowledges total
financial support from the Funda\c{c}\~ao de Amparo \`a Pesquisa
do Estado S\~ao Paulo (FAPESP). A.Z. and M.B. acknowledge support
of Ente Cassa di Risparmio di Firenze, Regione Toscana under
project CTOTUS, EU under ERA-NET CHIST-ERA project QSCALE, and
MIUR, under contract FIRB RBFR10M3SB.

\end{document}